\documentclass[twocolumn]{emulateapj}
\usepackage{amsmath}
\usepackage{graphicx}
\usepackage{natbib}
\newcommand{\ha}{H$\alpha$}   
\newcommand{\hb}{H$\beta$}

\newcommand{\kms}{\,km\,s$^{-1}$}  
\newcommand{\myr}{\,$M_{\sun}\,{\rm yr}^{-1}$}
\newcommand{\ro}{\,$R_{\sun}$}

\newcommand{\lo}{\,$L_{\sun}$}
\newcommand{\cmt}{\,cm$^{-3}$}

\newcommand{\es}{$\rm\,erg\,s^{-1}$}
\newcommand{\ecs}{$\rm\,erg\,cm^{-2}\,s^{-1}$}
\newcommand{\ecsa}{$\rm\,erg\,cm^{-2}\,s^{-1}\,\AA^{-1}$}
\slugcomment{Draft version March 19, 2018}

\shorttitle{Repeated transient jets from Z~And}
\shortauthors{Skopal et al.}

\begin{document}

\title{Repeated transient jets from a warped disk in 
       the symbiotic prototype Z~And: \\
       a link to the long-lasting active phase}

\author{Augustin Skopal\altaffilmark{1}}
\affil{Astronomical Institute, Slovak Academy of Sciences,
       059\,60 Tatransk\'{a} Lomnica, The Slovak Republic}
\author{Taya.~N.~Tarasova}
\affil{Scientific Research Institute, Crimean Astrophysical 
       Observatory, 298409 Nauchny, Crimea}
\author{Marek~Wolf}
\affil{Astronomical Institute, Charles University Prague, CZ-180\,00  
       Praha 8, V Hole\v{s}ovi\v{c}k\'ach 2, The Czech Republic}
\and
\author{Pavol~A.~Dubovsk\'y and Igor~Kudzej}
\affil{Vihorlat Astronomical Observatory, Mierov\'a 4, SK-066\,01
       Humenn\'e, The Slovak Republic}
\altaffiltext{1}{E-mail: skopal@ta3.sk}

\begin{abstract}
Active phases of some symbiotic binaries survive for a long time 
from years to decades. The accretion process onto a white dwarf (WD) 
sustaining long-lasting activity, and sometimes leading to collimated 
ejection, is not well understood. 
   We present the repeated emergence of 
highly collimated outflows (jets) from the symbiotic prototype 
Z~And during its 2008 and 2009-10 outbursts and suggest their 
link to the current long-lasting (from 2000) active phase. 
   We monitored Z~And with the high-resolution spectroscopy, 
multicolor $UBVR_{\rm C}$--and high-time-resolution--photometry. 
  The well-pronounced bipolar jets were ejected again during 
the 2009-10 outburst 
together with the simultaneous emergence of the rapid 
photometric variability ($\Delta m \approx 0.06$\,mag) on the 
timescale of hours, showing similar properties as those during 
the 2006 outburst. These phenomena and the measured disk-jets 
connection could be caused by the radiation-induced warping 
of the inner disk due to a significant increase of the burning 
WD luminosity. 
Ejection of transient jets by Z~And around outburst maxima 
signals a transient accretion at rates above the upper limit 
of the stable hydrogen burning on the WD surface. 
The enhanced accretion through the disk warping, supplemented 
by the accretion from the giant's wind, can keep a high luminosity 
of the WD for a long time, until depletion of the disk. 
In this way, the jets provide a link to long-lasting active 
phases of Z~And. 
\end{abstract}
\keywords{Stars: binaries: symbiotic --
          stars: individual: (Z~And) --
          ISM: jets and outflows}
\maketitle
%
%----------------------------------------------------------------------
%
\section{Introduction}
\label{s:intro}
Symbiotic stars are the widest interacting binaries with 
orbital periods of a few years. They consist of a red giant 
(RG) as the donor and a white dwarf (WD) as the accretor 
\citep[][]{boy67,kenyon86}. Symbiotic stars are well detached 
binary systems \citep[][]{ms99}, which implies that their 
activity is triggered via the wind mass transfer. 

According to the behavior of optical light curves (LC) we 
distinguish between the quiescent and active phases. 
During quiescent phases, the WD accretes throughout the accretion 
disk formed from the giant's wind. This process heats up the WD 
to $>10^5$\,K and increases its luminosity to 
$\sim 10^{1}-10^{4}$\lo, which in return ionizes 
a portion of the neutral wind from the giant, giving rise to 
the nebular emission \citep[][]{stb}. Optical LCs are 
characterized with a periodic wave-like variation. 
No sudden brightenings are indicated. 
On the other hand, active phases are characterized by a few 
magnitude (multiple) brightening(s)--outbursts--observed 
on the timescale of a few months to years/decades \citep[see, 
e.g., historical LCs of FN~Sgr, Z~And, AX~Per of][]{brandi+05,
l+f08,l+f13} with signatures of a mass outflow 
\citep[e.g.,][]{fc+95,esipov00,sok+06,mckeever11}. 
They are called `Z And-type' outbursts, because they were 
observed in the past for a prototype of the class of symbiotic 
stars -- Z~And \citep[][]{kenyon86}. 
In rare cases, these outbursts are followed by collimated 
ejection. To date, its signature in the optical spectra has 
been recorded only for a few objects: 
MWC~560 \citep[e.g.][]{tomov+90}, 
Hen~3-1341 \citep[][]{tomov+00,munari+05}, 
St\ha\,190 \citep[][]{munari+01}, 
Z~And \citep[][]{sk+pr06}, 
BF~Cyg \citep[][]{sk+13} and 
St\,2--22 \citep[][]{tomov+17}. 
%
%=======================|
%-- Fig. 1.: UBV LC  ---|
%=======================|
%
\begin{figure*}%[p!t]
\begin{center}
%
%\resizebox{\hsize}{!}{\includegraphics[angle=-90]{z_ubv_f1.eps}}
\resizebox{\hsize}{!}{\includegraphics[angle=-90]{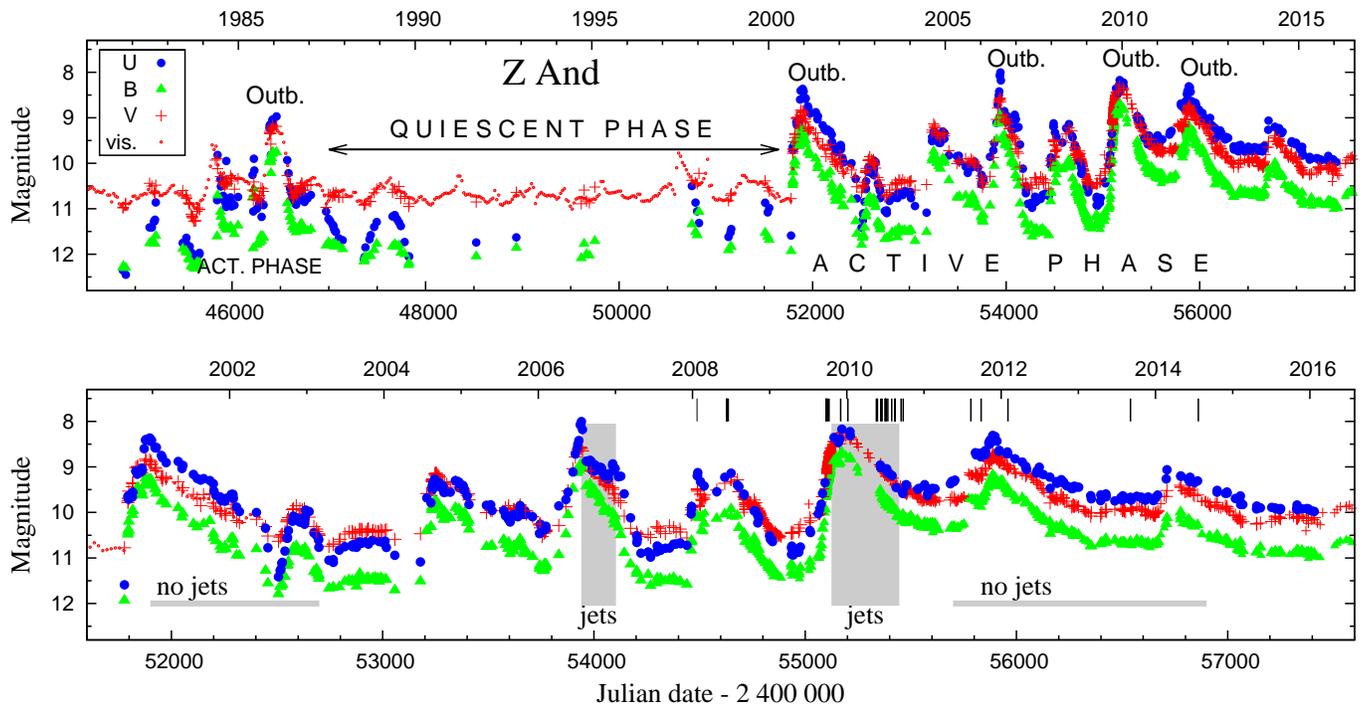}}
\end{center}
\caption{
Top: $UBV$ light curves of Z~And from 1981 covering its quiescent 
and active phases. References for the data used are in \cite{sk+00}. 
Bottom: Current active phase that started in September 2000 
showing multiple outbursts. Vertical bars indicate times of 
our spectra (Table~\ref{tab:spec}). The presence of bipolar 
jets is denoted by vertical gray belts. 
No jets were indicated during the major 2000-03 outburst 
\citep[][]{sk+06} and during 2011-12 and 2014 brightenings 
(Fig.~\ref{fig:haev}). 
Data are described in Sect.~\ref{ss:phot}. 
}
\label{fig:lc}
\end{figure*}

The Z~And binary consists of a late-type M4.5\,III RG and 
a WD accreting from the giant's wind on the 758-day orbit
\citep[e.g.][]{nv89,fekel+00}. 
During the quiescent phase, a high luminosity of its hot component, 
on the order of $10^3$\lo\ \citep[][]{muerset+91,sk05}, is 
generated by the stable nuclear burning of hydrogen on the 
WD surface \citep[e.g.,][]{pacrud80}. 
We call these systems nuclearly powered.\footnote{For a 
comparison, accretion-powered systems generate the luminosity 
of the order of $10^1$\lo\ by converting the gravitational 
potential energy of the WD via the accretion disk.} 
In September 2000, Z~And started a new active 
phase showing a series of outbursts with 
the main optical maxima in December 2000, July 2006, 
December 2009 and December 2011 (see Fig.~\ref{fig:lc}). 
At/after the 2006 maximum, highly collimated bipolar jets 
were detected for the first time as the satellite emission 
components to \ha\ and \hb\ lines \citep[][]{sk+pr06}. 
Their presence was confirmed by \cite{burmeister+07} and 
\cite{tomov+07}. 
Temporal development of jets and their possible origin 
were investigated by \cite{sk+09}. The event of jets was 
transient, being detected through the end of 2006, along the 
decrease of the star's brightness. During this period, 
a low amplitude irregular photometric variation from 
the low stage ($\la 0.02$\,mag) increased its amplitude 
to $\sim 0.06$\,mag on the timescale of hours. Its source 
was identified with a large disk around the WD. The authors 
ascribed these phenomena to a disruption of the inner parts 
of the disk caused by its radiation-induced warping. 

In this contribution, we present an analysis of collimated 
mass outflow, which we measured again during the 2008 and 2009--10 
outbursts. We compare the new jet event with that from 
the 2006 outburst to establish a more trustworthy disk-jets 
connection. 
In this way, we aim to understand better 
accretion process onto the WD that keeps the current active 
phase of Z~And for a long time, showing outbursts with or 
without jets. 
Our observations are described in Sect.~\ref{s:obs}. 
Their analysis and results are introduced in Sect.~\ref{s:anal}. 
A discussion of the results and a summary are provided in 
Sects.~\ref{s:dis} and \ref{s:sum}, respectively. 
\section{Observations and data reduction}
\label{s:obs}
Our observations of Z~And during its current active phase, 
from 2008 to 2014, were carried out at different 
observatories. 

\subsection{Spectroscopy}
\label{ss:spec}
(i) At the Crimean Astrophysical Observatory (denoted as CrAO 
in Table~\ref{tab:spec}) the high-resolution spectra were 
performed with the coud\'e spectrograph of the 2.6 m ZTSH telescope. 
The size of the ANDOR IKON-L CCD detector DZ936N was 
2048$\times$2048 pixels. The spectral resolving power was 
R$\sim$25\,000 around \ha. The low-resolution spectra 
(R$\sim$1\,000) were obtained using a spectrograph mounted at 
the Nasmyth telescope focus. % of the of 2.6 m ZTSH telescope. 
The detector was a SPEC-10 CCD camera (1340$\times$100 pixels). 
The spectrophotometric standard HR~8965 was observed each night 
to calibrate the measured flux in the star's spectrum. 
The primary reduction of the spectra, including the bias 
subtraction and flat fielding, was performed with the {\small SPERED} 
code developed by S. I. Sergeev at the CrAO. 

(ii)
At the Ond\v{r}ejov Observatory (Ondr.), spectra of Z~And were 
secured with a SITe-005 800$\times$2030 CCD detector attached 
to the medium 0.7-m camera of the coud\'{e} focus of the 
Ond\v{r}ejov 2.0 m telescope. The spectra were obtained 
between June 2010 and July 2014, and have a linear dispersion 
of 17.2\,\AA/mm and a 2-pixel resolving power of about 12\,600 
(11--12\,km\,s$^{-1}$ per pixel). Standard initial reduction of 
CCD spectra was carried out using modified {\small MIDAS} and 
{\small IRAF} packages. Final processing of the data was done 
with the aid of the {\small SPEFO}-package software 
\citep[][]{horn,skoda}. 

(iii)
At the David Dunlap Observatory, University of Toronto (DDO), 
spectra of Z~And were performed with a Jobin Yovon Horiba CCD 
detector (2048$\times$512 pixels of 13.5\,$\mu$m size) attached 
to the single dispersion slit spectrograph of the Cassegrain 
focus of the 1.88 m telescope. The slit width was 240\,$\mu$m 
corresponding to 1.5\,arcsec at the focal plane. Around the \ha\ 
region, the resolving power was $\sim$12\,000. 
Standard reduction of the spectra was performed with 
the {\small IRAF}-package software. 

A journal of the spectra is given in Table~\ref{tab:spec}. 
%
%   -----  Table 1 ----
%
\begin{table}%[p!t]
\centering
\caption{Log of spectroscopic observations}
\begin{tabular}{cccl}
\hline
Date       &    JD     & Region   & Obs.$^{\star}$ \\
(mm/dd/yyyy)& 2\,45... &   (nm)   &                \\
\hline
\multicolumn{4}{c}{2008 burst} \\
\hline
01/23/2008 & 4488.547  & 642-670  &  DDO \\
06/11/2008 & 4628.844  & 642-670  &  DDO \\
06/11/2008 & 4628.856  & 461-491  &  DDO \\
06/13/2008 & 4630.836  & 642-670  &  DDO \\
06/19/2008 & 4636.822  & 642-670  &  DDO \\
\hline
\multicolumn{4}{c}{2009-10 outburst} \\
\hline
09/26/2009 & 5100.535  & 335-740  &  CrAO$^a$ \\
10/01/2009 & 5106.237  & 653-659  &  CrAO$^b$ \\  
10/08/2009 & 5113.297  & 653-659  &  CrAO$^b$ \\  
12/02/2009 & 5168.235  & 653-659  &  CrAO$^b$ \\  
01/05/2010 & 5202.273  & 651-661  &  CrAO$^b$ \\  
05/21/2010 & 5337.499  & 353-757  &  CrAO$^a$ \\
05/25/2010 & 5342.465  & 653-661  &  CrAO$^b$ \\  
06/10/2010 & 5358.508  & 626-677  &  Ondr. \\   
06/16/2010 & 5364.561  & 626-677  &  Ondr. \\   
07/01/2010 & 5379.472  & 651-661  &  CrAO$^b$ \\  
07/02/2010 & 5380.503  & 622-674  &  Ondr. \\   
07/13/2010 & 5391.509  & 626-677  &  Ondr. \\   
08/01/2010 & 5410.481  & 651-661  &  CrAO$^b$ \\  
09/24/2010 & 5464.434  & 647-665  &  CrAO$^b$ \\
\hline
\multicolumn{4}{c}{2012 and 2014 brightenings} \\
\hline
08/10/2011 & 5784.379  & 639-691  &  Ondr. \\
09/28/2011 & 5833.270  & 640-691  &  Ondr. \\
02/02/2012 & 5960.221  & 640-691  &  Ondr. \\
09/04/2013 & 6540.445  & 641-688  &  Ondr. \\
07/22/2014 & 6861.495  & 641-688  &  Ondr. \\
\hline
\multicolumn{4}{l}{{\bf Notes:}~$^{\star}$Observatory, 
                  $^a$Nasmyth, $^b$coud\'e}
\end{tabular}
\label{tab:spec}
\end{table}
%--------------------------------------------------------------

\subsection{Photometry}
\label{ss:phot}
(iv) At the Skalnat\'e Pleso and Star\'a Lesn\'a (pavilion G2) 
observatories, classical photoelectric $UBVR_{\rm C}$ 
measurements were carried out by single-channel photometers 
mounted in the Cassegrain foci of 0.6 m reflectors. 
Data are plotted in Fig.~\ref{fig:lc}. Each value represents 
the average of the observations during a night. Corresponding 
inner uncertainties are of a few times 0.01\,mag in the $B$, 
$V$ and $R_{\rm C}$ bands, and up to 0.05\,mag in the $U$ band 
\citep[see][]{vanko+15a,vanko+15b}. The data were published by 
\cite{sk+12}. New observations (from November 2011) will be 
published elsewhere (Seker\'a\v{s} et al., in preparation). 
To get a better coverage of the investigated period, we 
complemented our photometry with the $BVR_{\rm C}$ CCD 
measurements available at the AAVSO database.
\footnote{https://www.aavso.org/data-download}. 

% Fast CCD photometry: 
(v) At the Star\'a Lesn\'a observatory (pavilion G1) the 
high-time resolution CCD photometry was performed during 
nights on 06/24/2007, 09/21/2007 and 07/11/2008. 
The {\small SBIG ST10 MXE} CCD camera (2184$\times$1472 pixels; 
6.8 $\mu$m) mounted at the Newtonian focus of the 0.5 m 
telescope was used \citep[see][ in detail]{pv05}. 
The star BD+47\,4192 
\citep[$V$ = 8.99, $B-V$ = 0.41, $U-B$ = 0.14, 
$V-R_{\rm C}$ = 0.19;][ and references therein]{sk+12} was used 
as the standard star for both photoelectric and CCD observations. 

(vi) At the Astronomical Observatory on the Kolonica Saddle, 
the fast CCD photometry was performed during the nights on 
11/01/2009, 11/02/2009 and 08/14/2011. 
A {\small FLI Pro\,Line PL1001E} CCD camera with the chip 
1024$\times$1024 pixels (24 $\mu$m) was attached to 
the Ritchey-Chretien telescope 300/2400 mm. 
The star TYC3641-00678-1 
\citep[B=9.43, V=9.05, Rc=8.83, Ic=8.62,][]{hh97} 
was used as the comparison. 
All CCD frames were dark subtracted, flat-fielded, and 
corrected for cosmic rays. 
Our high-time-resolution photometry is plotted 
in Fig.~\ref{fig:warp}. 

Arbitrary flux units of the high-resolution spectra around 
the \ha\ line were converted to absolute fluxes with the aid 
of $R_{\rm C}$ magnitudes corrected for the \ha\ equivalent 
width \citep[see Eq.~(10) of][]{sk07}. Magnitudes were 
converted to fluxes according to the calibration of 
\cite{hk82} and \cite{bessel79}. 
Observations were dereddened with $E_{\rm B-V}$ = 0.30 and
resulting parameters were scaled to a distance of 1.5\,kpc 
\citep[][]{muerset+91,mik+ken96}. 
Orbital phase was calculated according to the ephemeris of 
the inferior conjunction of the cool giant 
\citep[][]{fekel+00} as 
\begin{equation}
 JD_{\rm spec. conj.} = 
    2\,450\,260.2 + 759.0\times E .
\end{equation}

\section{Analysis and results}
\label{s:anal}
\subsection{Photometric evolution}
\label{ss:photevol}
Figure~\ref{fig:lc} shows the $UBV$ LCs of Z~And covering 
its quiescent and active phases from 1981. The recent active 
phase began in September 2000 \citep[][]{sk+00}. 
Our spectra (Table~\ref{tab:spec}) cover three brightenings 
in the recent evolution of the LC. The burst in 2008 with 
$U_{\rm max}\sim 9.2$, the main outburst that peaked at 
$U\sim 8.2$ on December 2009 and the 2012 and 2014 brightenings 
with $U_{\rm max}\sim 8.4$ and $\sim$9.0. 

To justify the connection between the rapid photometric variability 
and ejection of jets as suggested by \cite{sk+09}, we searched 
for a short-term photometric variability on the timescales of 
minutes to hours. During the 2007 quiescent phase, prior to 
the 2008 burst, we observed an irregular variation within 
$\Delta B \la 0.02$\,mag, comparable with that of standard 
stars. 
During the 2008 burst, we observed light variations with 
$\Delta B \sim 0.025$\,mag on the timescale of 1--2 hours, 
whereas at the maximum of the main 2009 outburst, the light 
variations increased to $\Delta B \sim 0.065$\,mag and enlarged 
its timescale to 7--9 hours throughout the whole night 
(Fig.~\ref{fig:warp}). 
Similar evolution in the rapid photometric variability was 
also indicated during the previous 2006 major outburst, when 
the jet features were measured for the first time 
\citep[see Fig.~3 of][]{sk+09}. 
%
%=====================================================|
%-- Fig. 2.: Jets during the 2008 burst: Ha and Hb ---|
%=====================================================|
%
\begin{figure}[p!t]
\begin{center}
\resizebox{\hsize}{!}{\includegraphics[angle=-90]{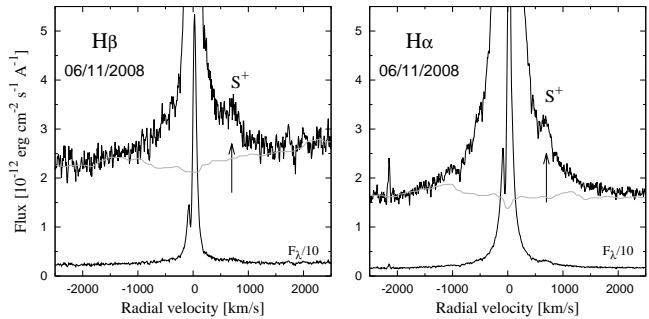}}
\end{center}
\caption{
Example of the \hb\ and \ha\ profile observed at the maximum 
of the 2008 burst. Only the S$^{+}$ satellite component was 
indicated at $\sim$700\kms. The gray line represents the 
model continuum of a normal M5 giant according to 
\cite{fluks+94}. 
}
\label{fig:hba08}
\end{figure}
%
%
%=================================================|
%-- Fig. 3.: Evolution in the H-alpha profile  ---|
%=================================================|
%
\begin{figure*}
%\centering
\begin{center}
\resizebox{\hsize}{!}{\includegraphics[angle=-90]{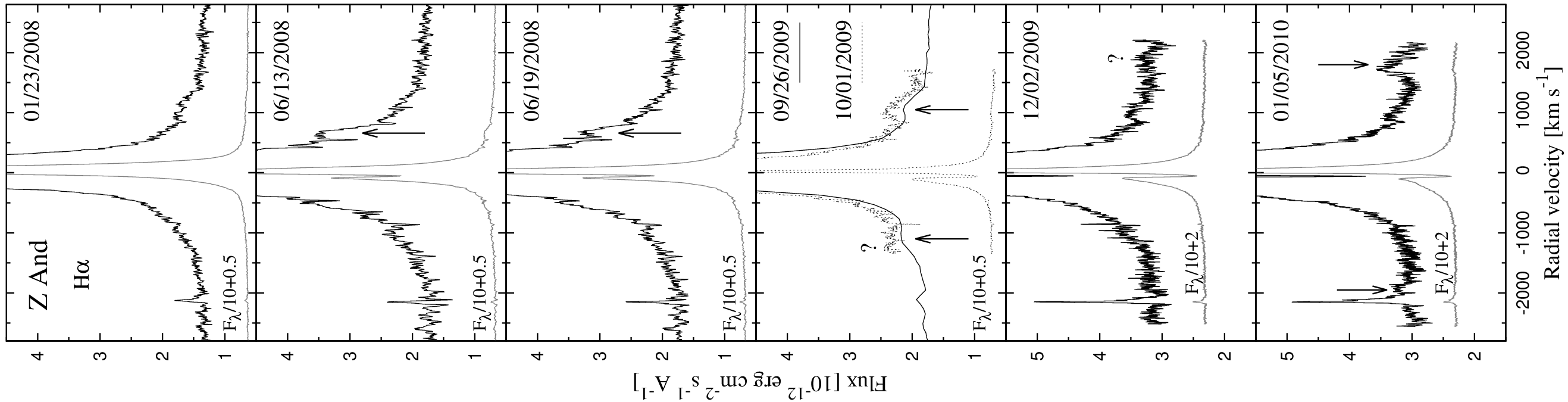}
                      \includegraphics[angle=-90]{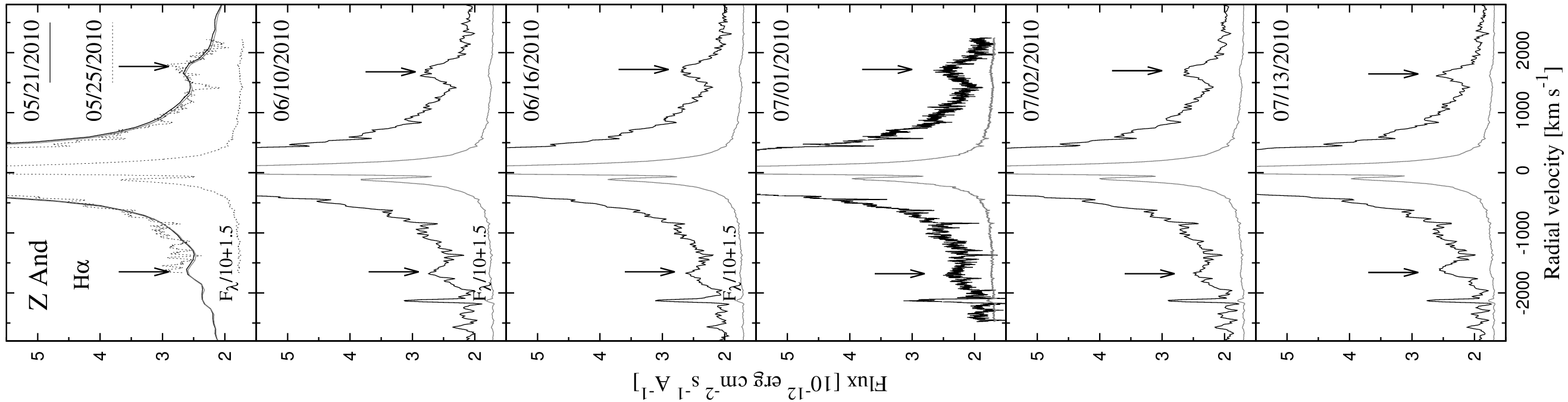}
                      \includegraphics[angle=-90]{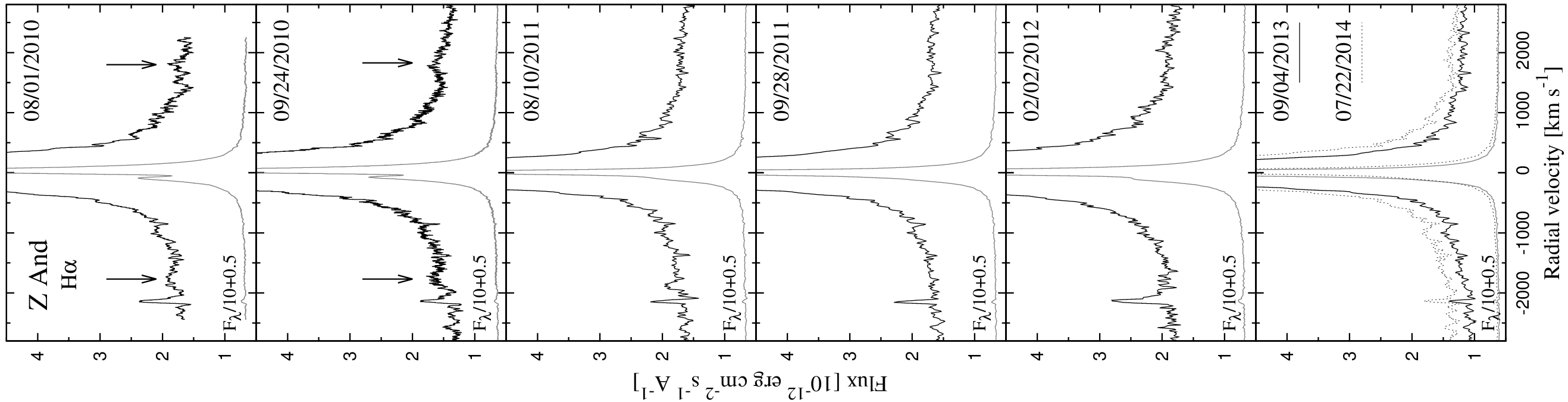}}
\end{center}
\caption{
Evolution of the \ha\ broad wings in our spectra 
(Sect.~\ref{ss:ha}). The satellite components are 
denoted by arrows. Their parameters are in Table~\ref{tab:par}. 
When the coverage of broad wings was insufficient, we compared 
low-resolution spectra (09/26/2009 and 05/21/2010). 
 The presence or absence of bipolar jets along the current 
active phase is denoted in Fig.~\ref{fig:lc}. 
}
\label{fig:haev}
\end{figure*} 
%
%
%============================================================|
%-- Fig. 4.: H-alpha broad wings & photometric variability --|
%============================================================|
%
\begin{figure*}[p!t]
\begin{center}
%
%\resizebox{\hsize}{!}{\includegraphics[angle=-90]{warping_f.eps}}
\resizebox{17cm}{!}{\includegraphics[angle=-90]{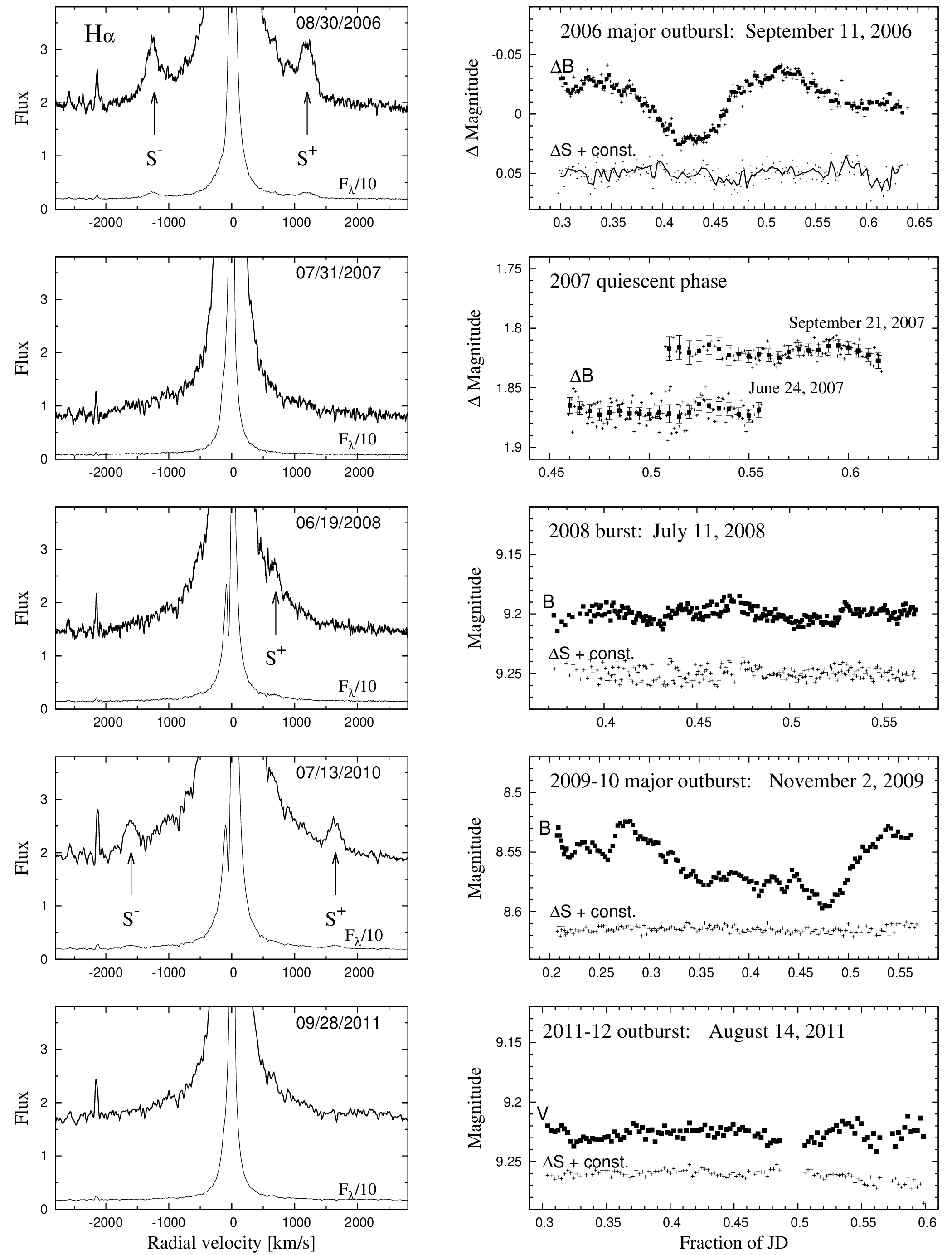}}
\end{center}
\caption{
\ha\ broad wings (left panels) and short-term variability in 
the optical continuum (right panels) at different stages of 
Z~And activity. During the 2006 and 2009-10 major outbursts, 
when the satellite S$^{-}$ and S$^{+}$ components emerged, 
a slow higher-amplitude photometric variation developed 
(Sect.~\ref{ss:photevol}). Data from 2006 and 2007 were 
published by \cite{sk+09}. Fluxes are in $10^{-12}$\ecsa. 
}
\label{fig:warp}
\end{figure*}
\subsection{Evolution of the \ha\ profile}
\label{ss:ha}
During the 2008 burst, the \ha\ line showed broad wings extended 
to $\sim \pm 1500$\kms\ and a sharp absorption cutting the 
emission core at $\sim$-50\kms. A faint $S^{+}$ satellite 
component was seen at $\sim$700\kms\ (Fig.~\ref{fig:haev}, 
spectra from January 23 to June 19, 2008). Its presence 
was confirmed by a similar feature in the \hb\ profile 
(Fig.~\ref{fig:hba08}). 

During the major 2009-10 outburst, the \ha\ profile was also 
of a P-Cyg type with broad wings (Fig.~\ref{fig:haev}, 
spectra from September 26, 2009 to September 24, 2010). 
The first satellite components were recorded on September 26, 2009, 
prior to the light maximum, at $\sim \pm1100$\kms. However, at 
the maximum of the star's brightness (our spectrum from 
12/02/2009) the satellite components were hardly recognizable. 
They appeared again when the brightness of Z~And began to fall 
(spectrum from January 5, 2010). 
Their position with respect to the reference wavelength increased 
to $\pm(1700 - 1800)$\kms\ during January to September 2010 
(Table~\ref{tab:par}, Fig.~\ref{fig:rvs}), when the star's 
brightness gradually decreased by $\sim$1 magnitude in $U$, 
from $\sim$8.2 to $\sim$9.2 (Fig.~\ref{fig:lc}). 

The last of our spectra taken around the following brightenings 
(December 2011 and January 2014; see Fig.~\ref{fig:lc}) did 
not show any clear signatures of satellite components in the 
\ha\ profile. We observed a simple emission core with 
a reduced wings and the width of the line with respect to 
the 2009-10 outburst (Fig.~\ref{fig:haev}). 

\subsection{Parameters of satellite components}
\label{ss:satel}

To determine measured parameters of the satellite components, 
we isolated them from the whole line profile by fitting 
the emission line core and its extended wings with two 
Gaussian functions. Then the residual satellite emissions 
were compared with additional Gaussians. Using their 
fitted parameters (the central wavelength, maximum, $I$, 
and the width $\sigma$) we determined the radial velocity 
of satellite components, $RV_{\rm S}$, their flux 
$F_{\rm S} = \sqrt{2 \pi}\,I\,\sigma$ and the width 
$FWHM_{\rm S} = 2\sqrt{2\ln(2)}\,\sigma$. 
Resolution of our spectra allowed us to estimate 
uncertainties in $RV_{\rm S}$ to 10$-$25\kms, in $F_{\rm S}$ 
fluxes within 10$-$20\% of the observed values 
and in $FWHM_{\rm S}$ widths within 0.2$-$0.3\,\AA. 
The results are given in Table~\ref{tab:par}. 
\subsection{Physical parameters of jets}
\label{ss:jets}

According to reasons discussed by \cite{sk+09}, the geometry 
of the emitting medium that gives rise to the satellite 
components can be approximated by a narrow cone with the peak 
at the central object characterized with an opening angle 
$\theta_0$. This implies that the measured narrow satellite 
components can be produced by highly collimated emitting 
particles -- jets. 
Therefore, using the measured parameters of the satellite 
components, we can determine some physical parameters of jets. 

%         Opening angle
%--------------------------------------
\subsubsection{Opening angle}
\label{sss:theta}
Assuming that jets were launched with a constant velocity, 
$v_{\rm jet}$, perpendicularly to the disk plane, which coincides 
with the orbital one (i.e. $RV_{\rm S} = v_{\rm jet}\cos(i)$), 
the opening angle can be approximated as 
\begin{equation}
\theta_0 = 2\,\sin^{-1}
   \Big[\frac{HWZI_{\rm S}}{RV_{\rm S}\tan(i)}\Big], 
\label{eq:theta}
\end{equation}
where $i$ is the orbital inclination \citep[see][]{sk+09}. 
Corresponding parameters from Table~\ref{tab:par}, orbital 
inclination $i = 63.\degr6 + 6.\degr9/-12.\degr2$ 
\citep[][]{shag17} and adopting $HWZI_{\rm S} = FWHM_{\rm S}$ 
yield the average value of the jet opening angle as 
\begin{eqnarray}
\theta_0 = 
   16.\degr5 +4.\degr7/-8.\degr3~~~{\rm for~~2008~burst}\hspace*{2.4cm}
\nonumber \\
              {\rm and} \hspace*{8.3cm}
\nonumber \\
\theta_0 = 
   9.\degr4 +2.\degr8/-4.\degr8~~~{\rm for~~2009-10~outburst},\hspace*{1.2cm}
\label{eq:nozzle}
\end{eqnarray}
after its maximum. Prior to the maximum of the 2009-10 outburst, 
$\theta_0 \sim 15.\degr2$. Uncertainties were determined 
as standard error of function (\ref{eq:theta}) using its 
total differential for the uncertainty in $i$. Influence of 
uncertainties in $FWHM_{\rm S}$ and $RV_{\rm S}$ 
(Sect.~\ref{ss:satel}) can be neglected. 
Individual values of $\theta_0$ are given in Table~\ref{tab:par}. 
Figure~\ref{fig:rvs} depicts also data from the 2006 outburst 
recalculated for $i = 63.\degr6$. 

%         Emission measure
%--------------------------------------
\subsubsection{Emission measure}
\label{sss:em}
Assuming that (i) the jet radiation is produced by the recombination 
transitions in the \ha\ line, (ii) the medium is completely ionized, 
(i.e. $n_{\rm e} = n_{\rm p} \equiv \bar{n}_{\rm jet}$) and 
radiates at a constant electron temperature $T_{\rm e}$, the 
luminosity of jets in \ha\ is related to the line emissivity, 
$\varepsilon_{\alpha} \bar{n}^2_{\rm jet}$, by
\begin{equation}
   L_{\rm jet}({\rm H\alpha}) = 
   \varepsilon_{\alpha}\,\bar{n}^2_{\rm jet}\,V_{\rm jet}, 
\label{eq:lha}
\end{equation}
%----------------------------------------------------
where $\varepsilon_{\alpha}$ is the volume emission coefficient 
in \ha, $\bar{n}_{\rm jet}$ is the mean particle concentration 
and $V_{\rm jet}$ is the volume of the jets. 
For the optically thin medium of jets, the luminosity can be
determined from the observed fluxes as    
$L_{\rm jet} = 4\pi d^2 \times F_{\rm S}$.
According to the definition of the emission measure, 
$EM_{\rm jet} = \bar{n}^2_{\rm jet}\,V_{\rm jet}$, and 
assumptions above, we can express it as, 
\begin{equation}
   EM_{\rm jet}({\rm H\alpha}) = 
           4 \pi d^2 \frac{F_{\rm S}}{\varepsilon_{\alpha}}.
\label{eq:em}
\end{equation}
%----------------------------------------------------
The emission measure for one jet is of a few $\times 10^{57}$\cmt\ 
for $\varepsilon_{\alpha}(T_{\rm e} = 2\times 10^{4}$\,K) = 
1.83$\times 10^{-25}\rm erg\,cm^{3}\,s^{-1}$ 
\citep[e.g.][]{ost89} (Table~\ref{tab:par}). 
%
%-------------------- Table 2 -----------------------
%
\begin{table*}[p!t]
\centering
\caption{Parameters of Gaussian fits to the satellite emission 
  components $S^-$ and $S^+$: radial velocity $RV_{\rm S}$ 
  (\kms), flux $F_{\rm S}$ (10$^{-12}$\ecs) and 
  $FWHM_{\rm S}$ (\kms). 
  Derived parameters are: opening angle $\theta_0$ ($\degr$) 
  and emission measure of both jets, EM$_{\rm jet}$ 
  (10$^{57}$\cmt) 
}
\begin{tabular}{cccccccccc}
\hline
Date & 
\multicolumn{2}{c}{$RV_{\rm S}$} &
\multicolumn{2}{c}{$F_{\rm S}$} &
\multicolumn{2}{c}{$FWHM_{\rm S}$} &
\multicolumn{2}{c}{$\theta_0$} &
EM$_{\rm jet}$ \\
%\multicolumn{2}{c}{$\dot M_{\rm jet}$} \\
%
\hline
 mm/dd/yyyy & S$^-$ & S$^+$ & S$^-$& S$^+$ & S$^-$& S$^+$& S$^-$& S$^+$ \\
\hline
%               RVs   RVs   Fs   Fs   FW   FW    O     O    EM
 06/11/2008 &  -   & 707 & -  &2.3  &  -  &194 & -   &15.7&3.4   \\
 06/13/2008 &  -   & 661 & -  &3.0  &  -  &215 & -   &18.6&4.4   \\
 06/19/2008 &  -   & 693 & -  &1.7  &  -  &183 & -   &15.1&2.5   \\
 09/26/2009$^{\star}$ & -1157& 1105&    &     &     &   &  &  &  \\
 10/01/2009$^{\dagger}$ &     & 1060&   & 2.3 &     &290&  &15.6~& \\
 10/08/2009$^{\dagger}$ &     & 1040&   & 2.2 &     &269&  &14.8~& \\
 12/02/2009 & \multicolumn{9}{c}{no ~ jets} \\
 01/05/2010 & -2000& 1790& 3.3& 5.1 & 301 &301 &8.57 &9.58 &12.4 \\
 05/21/2010$^{\star}$ & -1670 &1740&    &     &     &   &  &  &  \\
 05/25/2010$^{\dagger}$ &     & 1750&   & 5.4 &     &312&  &10.1~& \\
 06/10/2010 & -1714& 1698& 4.0& 4.8 & 274 &282 &9.10 &9.45 &12.9 \\
 06/16/2010 & -1723& 1721& 3.7& 4.2 & 296 &290 &9.78 &9.60 &11.6 \\
 07/01/2010 & -1723& 1720& 3.1& 3.7 & 301 &290 &9.95 &9.60 &10.0 \\
 07/02/2010 & -1723& 1703& 3.1& 3.9 & 290 &280 &9.59 &9.36 &10.3 \\
 07/13/2010 & -1625& 1639& 3.1& 3.1 & 247 &247 &8.65 &8.58 & 9.1 \\
 08/01/2010 & -1750& 1750& 1.8& 2.0 & 301 &301 &9.80 &9.80 & 5.6 \\
 09/24/2010 & -1755& 1830& 1.3& 1.3 & 280 &290 &9.10 &9.02 & 5.3 \\
\hline
\multicolumn{10}{l}{{\bf Notes:}~$^{\star}$low-resolution spectrum, 
~$^{\dagger}$spectrum does not cover the $S^-$ component}
\end{tabular}
\label{tab:par}
\end{table*}
%
%         Radius of jets
%--------------------------------------
\subsubsection{Radius}
\label{sss:radius}
The conical shape of jets defines their volume as 
$V_{\rm jet} = 1/3 R^3_{\rm jet}\times\Delta\Omega$, where 
$\Delta\Omega = 2\pi [1 - \cos(\theta_0/2)]$ is the solid 
angle of the jet nozzle. Using Eq.~(\ref{eq:lha}) the jet 
radius, $R_{\rm jet}$, can be expressed by means of the 
parameters $L_{\rm jet}$ and $\Delta\Omega$ as a function 
of $\bar{n}_{\rm jet}$, 
\begin{equation}
 R_{\rm jet} = \left(\frac{3 L_{\rm jet}({\rm H\alpha})}
                          {\varepsilon_{\alpha}(H,T_{\rm e})
                           \Delta\Omega}\, 
                           \bar{n}_{\rm jet}^{-2} \right)^{1/3}.
\label{eq:rjet1}
\end{equation}
For average values of $F_{\rm S}$ and $\theta_0$ of one jet 
measured during 2010 
(Table~\ref{tab:par}, Eq.~(\ref{eq:nozzle})), the average jet 
radius can be expressed as a function of $\bar{n}_{\rm jet}$ 
as 
\begin{equation}
  \log(R_{\rm jet}/{\rm AU}) = 
      6.8 - \frac{2}{3}\log(\bar{n}_{\rm jet}) .
\label{eq:rjet2}
\end{equation}
%
%--------------------------------------
%       Mass-loss rate
%--------------------------------------
\subsubsection{Mass-loss rate}
\label{sss:dotm}
According to the mass continuity equation, the geometrical 
and kinematics parameters of the jets allow us to determine 
the mass-loss rate through jets as 
%-------------------------------------------
\begin{equation}
 \dot M_{\rm jet} = \Delta\Omega\,R^2_{\rm jet}\,\mu m_{\rm H}\,
                  \bar{n}_{\rm jet}v_{\rm jet}, 
\label{eq:dotm1}
\end{equation}
%-------------------------------------------
where $\mu$ is the mean molecular weight and $m_{\rm H}$ is
the mass of the hydrogen atom. 
For the average value of the jet velocity during 2010, 
$v_{\rm jet} = RV_{\rm S}/cos(i) = 3900$\kms\ 
(Table~\ref{tab:par}) and the angle of the jet nozzle 
of 9.\degr4, we obtain 
%Eq.~(\ref{eq:dotm}) has a form, 
%
\begin{equation}
  \log(\dot{M}_{\rm jet}/M_{\odot}{\rm yr^{-1}}) = -2.6 - 
                \frac{1}{3}\log(\bar{n}_{\rm jet})
\label{eq:dotm2}
\end{equation}
and/or as a function of the radius $R_{\rm jet}$ as 
\begin{equation}
  \log(\dot{M}_{\rm jet}/M_{\odot}{\rm yr^{-1}}) = 
      -6.0 + \frac{1}{2}\log(R_{\rm jet}/{\rm AU})
\label{eq:dotm3}
\end{equation}
for one jet. 
Determination of $\dot M_{\rm jet}$ thus requires a knowledge 
of $\bar{n}_{\rm jet}$ or $R_{\rm jet}$ 
\citep[see][ for details]{sk+09}. 
%
%===============================================|
%-- Fig. 5.: SED in outburst + decretion disk --|
%===============================================|
%
\begin{figure*}%[p!t]
\begin{center}
\resizebox{16cm}{!}{\includegraphics[angle=-90]{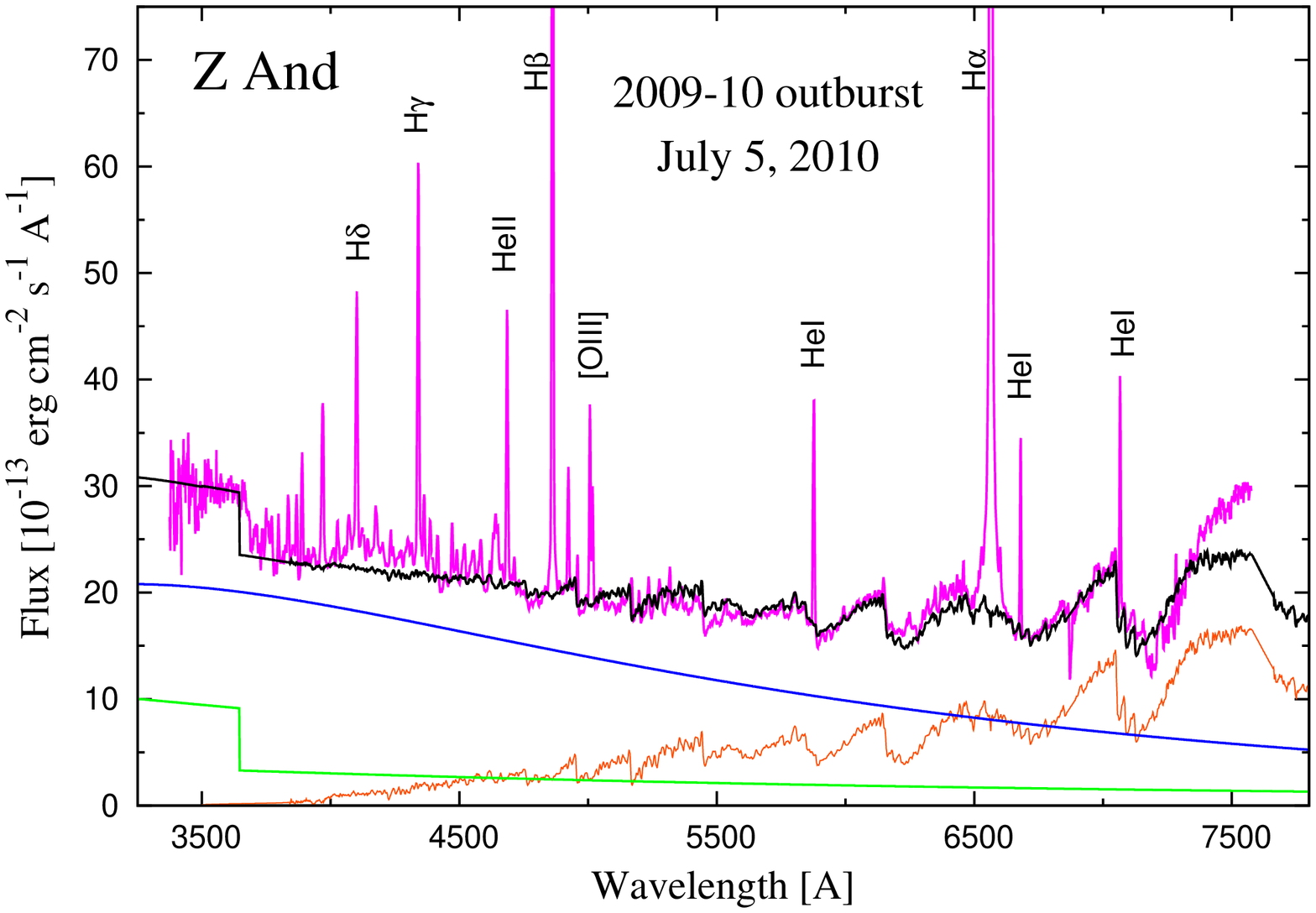}
\hspace*{1cm}
            \includegraphics[width=14cm,angle=-90]{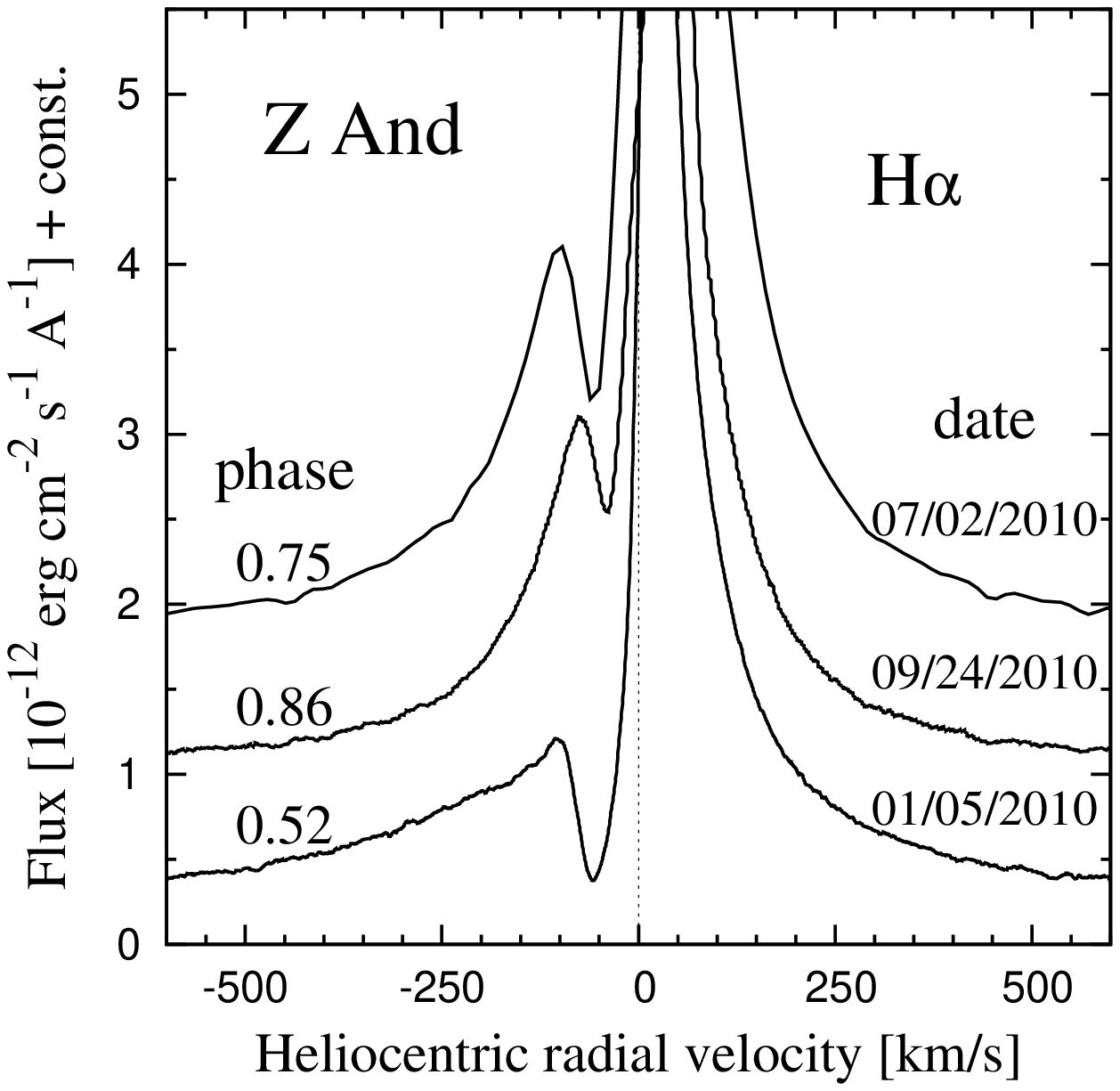}}
\end{center}
\caption{
Left: Optical spectrum from the 2009-10 outburst (magenta line) 
and its model SED (solid black line). It is given by superposition 
of radiation from the warm pseudophotosphere (blue line), 
the cool giant (orange line) and the nebula (green line). 
Right: Absorption component in the P-Cyg profile of the \ha\ 
line during this outburst indicates expansion of the warm 
pseudophotosphere. 
}
\label{fig:sed}
\end{figure*}
%
%
%==========================|
%-- Fig. 6.: RVs of jets --|
%==========================|
%
\begin{figure}%[p!t]
\begin{center}
%
%\resizebox{8cm}{!}{\includegraphics[angle=-90]{rvs_theta_lc.eps}}
\resizebox{\hsize}{!}{\includegraphics[angle=-90]{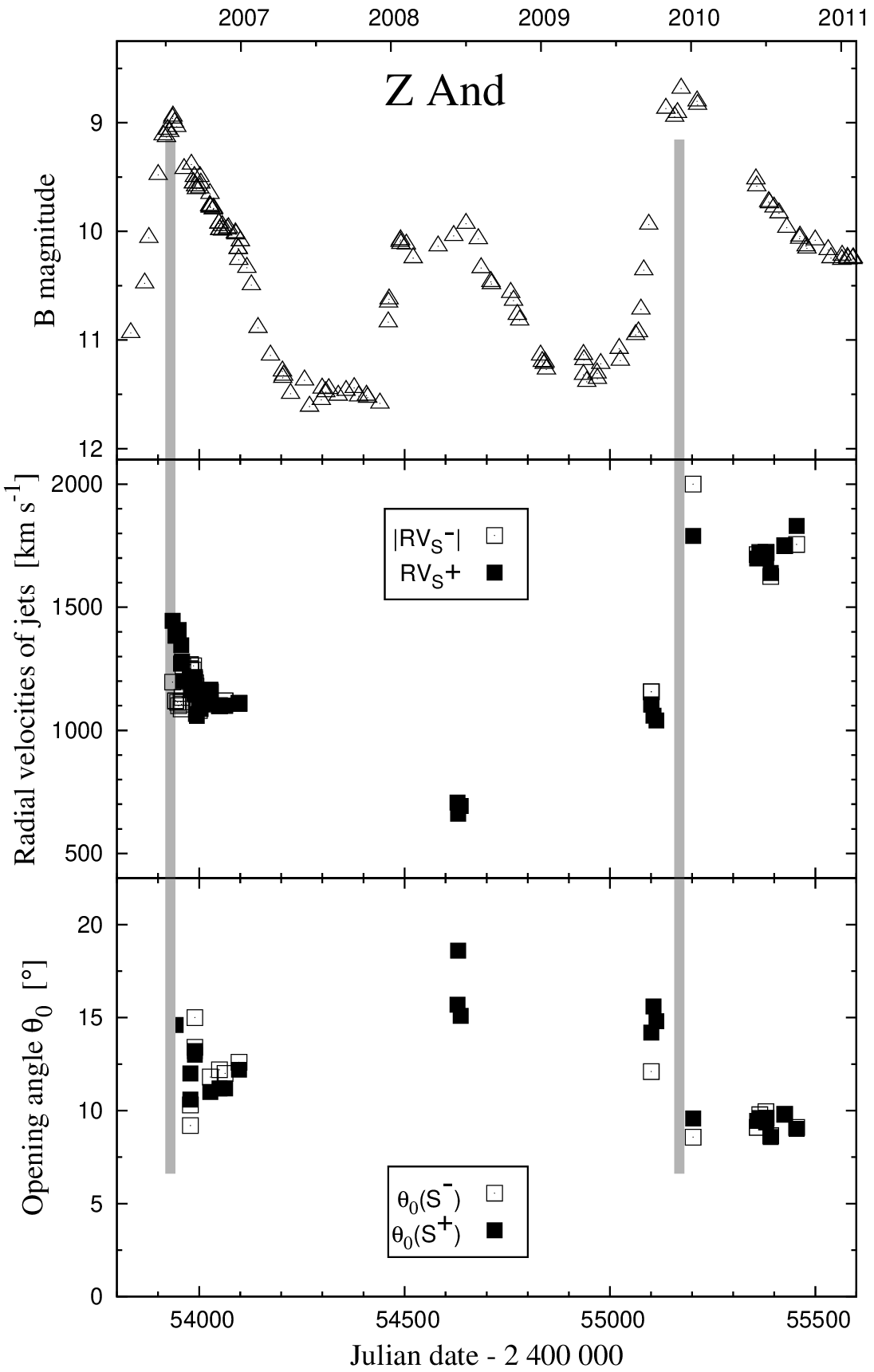}}
\end{center}
\caption{
Evolution of $RV_{\rm S}$ (middle) and $\theta_0$ (bottom) 
along the $B$-LC (top) of the 2006, 2008 and 2009-10 outbursts. 
Maxima of 2006 and 2009-10 events are denoted by vertical bars. 
Uncertainties in $RV_{\rm S}$ and magnitudes are within 
the size of points, and those in $\theta_0$ are described in 
Sect.~\ref{sss:theta}. 
Data are from \cite{burmeister+07}, \cite{tomov+07}, 
\cite{sk+09} and our Table~\ref{tab:par}. 
}
\label{fig:rvs}
\end{figure}
%
%------------------------------------------
% Comparison of 2006 and 2009-10 outbursts
%------------------------------------------
\subsection{Comparison with the 2006 outburst}
\label{ss:comp}
%
%...a two-sided jet. 
To date, bipolar jets in the form of satellite components to 
the \ha\ (\hb) line were identified in the spectrum of Z~And 
only during the 2006 and 2009-10 outbursts (Fig.~\ref{fig:lc}). 
Our monitoring of the latter showed that the spectroscopic 
and photometric characteristics of both outbursts are very 
similar. We summarize them as follows: 

(i) 
The jets appeared around their optical maxima, and were gradually 
vanishing along the decrease of the star's brightness. Thus the 
event of jets was transient (Figs.~\ref{fig:rvs} and 
\ref{fig:djc}). 

(ii)
In both cases, large variation in $RV_{\rm S}$ were observed 
at the beginning of their emergence, around optical maxima, 
and then settled on a constant level (see Fig.~2 of \cite{tomov+07}, 
Fig.~6 of \cite{sk+09} and Fig.~\ref{fig:rvs} here). 

(iii) 
During both outbursts, rapid light variation on the timescale 
of hours within $\Delta m \sim 0.06$\,mag developed from 
irregular $\la$0.02\,mag fluctuations outside the outbursts. 
(see Fig.~3 in \cite{sk+09} and Fig.~\ref{fig:warp} here). 

The outbursts diversified only in duration of the star's 
brightness decline with jets, and in $RV_{\rm S}$. 
For the 2009-10 outburst, the period with jets lasted 
for $\approx$10 months and satellite components settled 
on $\pm(1700 - 1800)$\kms, whereas for the 2006 outburst 
the relevant quantities were $\approx$5 months and 
$\sim\pm1100$\kms, respectively (Figs.~\ref{fig:rvs} and 
\ref{fig:djc}). 

\section{Discussion}
\label{s:dis}
Here we discuss basic conditions of the jet's ejection by the 
nuclearly powered WD in the symbiotic binary Z~And, and 
its possible consequences for the accretion process during 
Z~And-type outbursts of symbiotic stars. 

\cite{sk+09} interpreted the ejection of jets during the 2006 
outburst as a result of disruption of the inner parts of the 
disk due to the radiation-induced warping caused by 
a significant increase of the WD luminosity at the outburst 
maximum. 
Close similarity of both the outbursts with jets 
(Sect.~\ref{ss:comp}) suggests their common nature. 

However, there is no theoretical application of the 
radiation-induced warping of disks around WDs in symbiotic 
binaries. Therefore, we first briefly introduce main results 
of its application to selected types of objects. 

\subsection{On the radiation-induced warping of disks}
\label{ss:disdis}
\cite{petter77} found that the effect caused by the pressure 
of radiation from the center of an X-ray source on the disk 
structure is very important. 
\cite{iping+pett90} confirmed that radiation forces can make 
rings of the disk precess in either direction, and change 
their inclination angle. In this way, they explained the 
inclination of the disk in Her~X-1.\footnote{Her~X-1 is an 
intermediate-mass X-ray binary containing an accreting 
neutron star with jets \citep[][]{eijnden+18}.}
Using a simple analytic approach, \cite{p96} showed that 
even an initially flat disk is unstable to warping, because 
the surface of a warped disk is illuminated by a central 
radiation source in a non-uniform manner. 
\cite{lp96} showed that also the accretion disk around the 
central stars of PNe can become unstable to a radiation-induced 
self-warping. \cite{lp97} demonstrated that this effect is 
accompanied by a wobbling motion in both the inclination and 
azimuthal directions. They used the term "wobble" to describe 
rather erratic motions. 
\cite{p97} investigated the effect of radiation-induced warping 
on accretion disks around massive black holes in active 
galactic nuclei. %Assuming that jets emanate perpendicularly 
%to the local surface of the inner disk, 
He calculated that the axis of the jets can be severely 
misaligned from the normal to the outer disk. 
\cite{southwell+97} derived radiation-induced precession of 
the disk with a timescale of the order of months for a strong 
supersoft X-ray source CAL~83. 
\cite{w+p99} explored the effect of self-induced warping of 
disks in different types of X-ray binaries. They found that 
at high luminosities the inner disk can tilt through more than 
90$\degr$, which may explain the torque reversals in systems 
such as 4U~1626-67.\footnote{4U~1626-67 is an ultra-compact 
X-ray binary bearing a neutron star with the orbital period 
of 42\,min \citep[][ and references therein]{beri+18}.}

For symbiotic stars, a disruption of the disk has been considered 
for the accretion-powered symbiotic system CH~Cyg by \cite{s+k03}. 
They interpreted the change of the fastest variations (timescale 
of minutes) into smooth, hour-timescale variations by disruption 
of the inner disk in association with the mass ejection event. 
For the nuclearly powered symbiotic star Z~And, \cite{sk+09} 
suggested that the jets ejection during its 2006 outburst could 
be also triggered by the radiation-induced disk warping. 

In the following section, we will show how the basic conditions 
for the disk warping and jet ejection are supported by 
observations for the Z~And outbursts with jets. 
\subsection{Indication of radiation-induced warping of the disk 
            during Z~And outbursts with jets}
\label{ss:warping}
According to Sect.~\ref{ss:disdis} the necessary ingredients 
for disk warping and jets ejection are: 
(i) The presence of a disk, 
(ii) emergence of a strong central radiation source, and
(iii) an observational response of the wobbling motions 
      of the disk. 
The corresponding critical results confirming these terms for 
the case of Z~And are described as follows. 

(i) 
The presence of a disk-like formation around the WD during active
phases of symbiotic stars was proven by modeling their UV to 
near-IR SED \citep[see][]{sk05}. The disk-like structure
is indicated by the two-temperature-type of the hot component
spectrum. The cooler spectrum is produced by a warm stellar 
pseudophotosphere radiating at $1-2\times 10^{4}$\,K, whereas 
the hotter one is represented by the highly ionized emission 
lines and a strong nebular continuum. 
The former is not capable of producing the observed nebular
emission and thus the latter signals the presence of a strong 
ionizing source ($\gtrsim 10^{5}$\,K) in the system, which 
is not seen directly by the outer observer.
This type of the SED can be explained by a disk-like structure 
of the hot component viewed under a high inclination angle. Then 
the flared outer rim of the disk (which is the warm pseudophotosphere 
indicated by model SED) occults the central ionizing source in 
the line of sight, while the nebula above/below the disk can 
easily be ionized \citep[see Fig.~27 of][]{sk05}. 

During the 2009-10 outburst of Z~And, creation of such a 
disk around the burning WD was documented by models of SED from 
September 2009 to November 2010 \citep[see Fig.~8 of][]{t+s12}. 
Here, Fig.~\ref{fig:sed} shows example of a representative model 
SED of the optical spectrum from July 5, 2010.  
The two-temperature-type of the hot component spectrum consists 
of a stellar radiation produced by a warm pseudophotosphere 
(the cooler component) and a strong nebular emission (the hotter 
component). The former radiates at $\sim 9000$\,K and has the 
luminosity of $\sim 700$\lo\ (the blue line in the figure), while 
the latter composes of the nebular continuum with a high emission 
measure $EM \sim 1.7\times 10^{60}$\cmt\ (the green line) and 
emission lines of highly ionized elements (e.g. H\,{\small I}, 
He\,{\small I}, He\,{\small II}, [O\,{\small III}]). 
The warm pseudophotosphere generates the flux of hydrogen ionizing 
photons $L_{\rm H}\sim 3\times 10^{42}$\,s$^{-1}$, which is not 
capable of producing the observed $EM$ that requires 
$L_{\rm H} = \alpha_{\rm B}({\rm H},T_{\rm e})\times EM 
             \sim 2\times 10^{47}$\,s$^{-1}$ 
for the total hydrogen recombination coefficient 
$\alpha_{\rm B}({\rm H},T_{\rm e}) = 1.0\times 10^{-13}$ 
$\rm cm^{3}\,s^{-1}$ \citep[e.g.,][]{nv87}. 
This implies that the warm pseudophotosphere cannot be a sphere, 
but a disk, as described above. 
Independently, the presence of jets indicates the presence of 
a disk, because jets require an accretion disk 
\citep[e.g.][]{livio97}. 

(ii) 
The emergence of a strong central radiation source 
is connected with outbursts, when a significant increase of 
the central source luminosity is indicated. 
For example, the emission measure of $1.7\times 10^{60}$\cmt\ 
requires the luminosity of the ionizing source to be of 
the order of $10^{37}$\es\ \citep[see][ in detail]{sk+17}. 
 Also a high $EM \sim 9.4\times 10^{60}$\cmt, as indicated 
around the maximum of the 2006 outburst 
\citep[see Sect.~3.1. of][]{sk+09}, suggests similarly high 
luminosity. 
For the 2000-03 outburst, \cite{sok+06} estimated the hot star 
luminosity to $10^{4}$\lo, lasting approximately for one year. 
Thus the high luminosity of the central source during outbursts 
can illuminate the disk, which can become unstable to warping. 

(iii)
 Observational response of the radiation-induced warping 
of the disk can be associated with the higher-amplitude 
photometric variability ($\Delta m\sim 0.06$\,mag) on the 
timescale of hours that emerges exclusively during major 
outbursts, during which the bipolar jets are launched 
(Fig.~\ref{fig:warp}). 
According to models SED, this type of variability is produced 
by the warm pseudophotosphere, i.e., the outer rim of the disk, 
because its contribution dominates the $B,V$ passbands 
(see Fig.~\ref{fig:sed} and point (i) above).  
In this way, we directly indicate the response of the disk warping 
in the form of the short-term variations in the continuum. 
As the dynamical timescale of the disk is comparable with the 
timescale of the smooth light variations (see Appendix~\ref{app1}), 
this can be caused by a variable projection of the disk surface 
into the line of sight. We thus observe wobbling of the outer 
parts of the disk -- in agreement with the general view that the 
warped disk starts to wobble or precess 
\citep[][ Sect.~\ref{ss:disdis}]{lp97}. 
Alternatively, the smooth light variations could be partly 
caused by the brightness variation of the outer disk due to 
reprocessing flickering light from its inner warping part 
(see Appendix~\ref{app2}). 
\subsection{Jets' radial velocities from the warping disk}
\label{ss:rvs}
Complex variations in $RV_{\rm S}$ around the maxima of both 
2006 and 2009-10 outbursts (Sect.~\ref{ss:comp}) could also 
be a direct result of the radiation-induced disk warping. 
Assuming jets to be 
perpendicular to the inner disk, their initial chaotic wobbling 
motions can produce shifts in $RV_{\rm S}$ due to the different 
inclination of a jet to the line of sight, as considered 
for PNe by \cite{lp97}. 
One can also imagine that the twisted inner disk can cause 
that axis of both jet nozzles not to be parallel, which causes 
the observed jet's asymmetry \citep[see Fig.~6 of][]{sk+09}. 

However, after a short time of $\sim$1 and a few months, the 
values of $RV_{\rm S}$ settled at around $\pm\,1100$\kms\ 
and $\pm$(1700 -- 1800)\kms\ for 2006 and 2009-10 outburst, 
respectively. During the small 2008 burst, the only present S$^{+}$ 
component was located at $\sim$700\kms\ (Table~\ref{tab:par}). 
This suggests that the wobbling motions of the disk ceased. 

Diversity of the final $RV_{\rm S}$ during each outburst could 
be caused by different conditions for the radiation-driven 
warping during these events. 
According to \cite{p96}, the radiation-driven warping of the 
disk can occur at radii, which depend on the central star mass 
and luminosity, accretion rate, and viscosities in the disk. 
During the Z~And-type outbursts, additional energy is liberated 
by thermonuclear burning on the WD surface due to a transient 
high accretion rate \citep[see, e.g.,][ for AG~Peg]{sk+17}. 
Depending on its quantity during different outbursts with 
different properties of the disk, the warping can occur at 
different radii. Qualitatively, the material liberated at 
different heights above the WD then converts different 
amounts of the gravitational potential energy 
into the kinetic energy at its accretion and thus can be 
expelled at different velocities in the form of jets. 
The energy power of these outbursts, as given by their brightness 
maximum and duration, seems to be consistent with this view 
(see Fig.~\ref{fig:lc}). 
\subsection{Disk-jets connection in Z~And}
\label{ss:djc}
\begin{figure}[p!t]
\begin{center}
\resizebox{\hsize}{!}{\includegraphics[angle=-90]{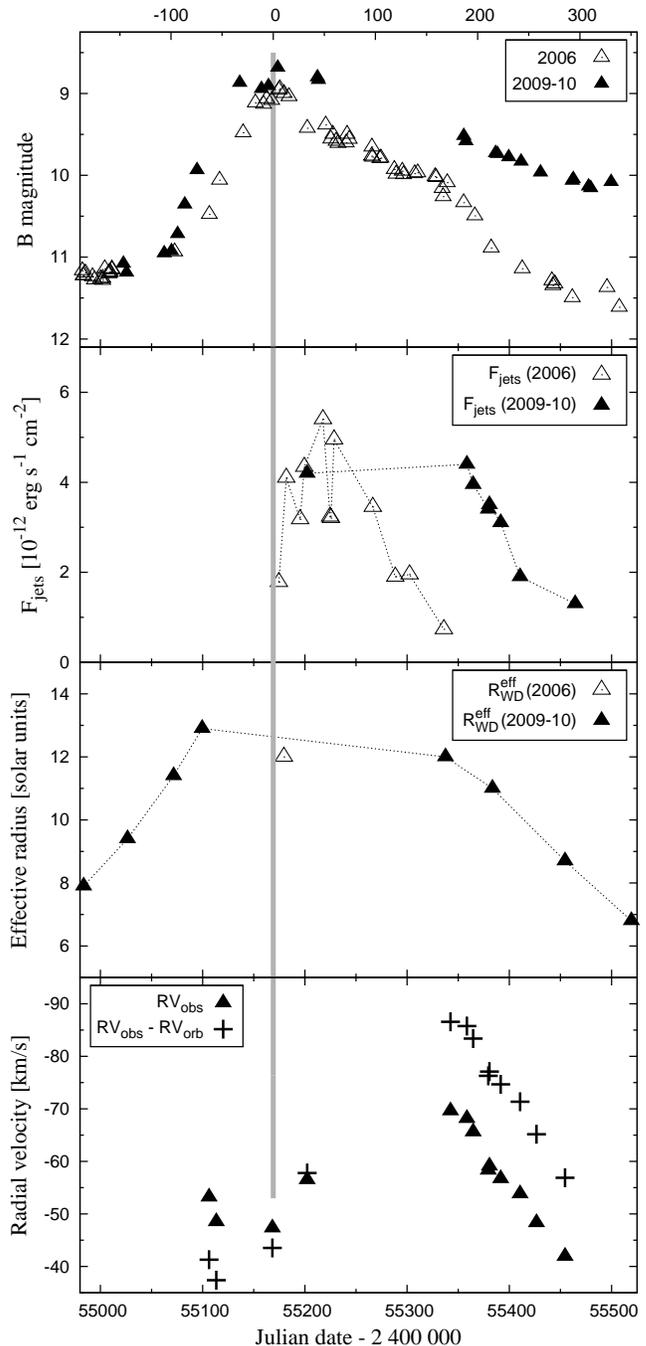}}
\end{center}
\caption{
Disk-jets connection during the 2006 and 2009-10 outbursts. 
The top panel compares their $B$-LCs shifted to their maxima 
(vertical bar). 
The second panel shows evolution of average fluxes of jets, 
$F_{\rm jets} = (F_{\rm S^{+}} + F_{\rm S^{-}})/2$ 
\citep[data from][ and our Table~2]{tomov+07,sk+09}. 
The third panel plots the evolution of the effective radius, 
$R_{\rm WD}^{\rm eff}$, of the warm pseudophotosphere, 
and the bottom panel displays radial velocities of the 
absorption component in the \ha\ profile (crosses are values 
corrected for the orbital motion of the hot component 
according to elements of \cite{fekel+00}). 
%See Sect.~\ref{ss:djc}. 
}
\label{fig:djc}
\end{figure}

Figure~\ref{fig:djc} shows the evolution of the measured flux 
of jets and the effective radius of the warm pseudophotosphere, 
$R^{\rm eff}_{\rm WD}$, along the 2006 and 2009-10 outbursts. 
Note that the disk radius $R_{\rm D} \propto R^{\rm eff}_{\rm WD}$ 
\citep[e.g., Eq.~(11) of][]{sk+11}. Effective radii were derived 
from modeling the SED during the 2009-10 outburst 
\citep[see Table~5 of][]{t+s12}. From the 2006 outburst, there 
is only one estimate of $R^{\rm eff}_{\rm WD}$ available 
around the optical maximum \citep[][]{sk+09}. 
Gradual decrease of the disk radius and the flux of jets along 
the decline of the star's brightness implies a gradual dilution 
of the disk. 

Shrinking of the disk radius is also indicated by the radial 
velocities of the absorption component in the \ha\ profile 
that decelerate along the decline of the star's brightness 
(bottom panel of Fig.~\ref{fig:djc}). 
This effect can be a result of expansion of the outer parts of 
the disk leading to a shrinkage of the optically thick part of 
the warm pseudophotosphere. Therefore, the optically thick wind 
in the \ha\ line at/above the warm pseudophotosphere will also 
originate at smaller radii, where it is driven outside with 
a smaller velocity. 

Below we describe how the disk--jets connection can be 
understood in the context of the accretion process onto the WD 
during Z~And outbursts with jets.
%
% !Therefore, emergence of the smooth light variation may indicate 
% !the warping of the inner disk that gives rise to collimated mass 
% !ejection as proposed in Sect.~\ref{ss:warping}. 
%
%===================================|
%-- Fig. 7.: Disk-jets connection --|
%===================================|
%
\subsubsection{On the accretion during Z~And outbursts with jets}
\label{ss:djcexpl}
Significant increase of the WD luminosity around the optical 
maximum, the indication of the disk warping 
(Sects.~\ref{ss:warping} and \ref{ss:rvs}), and the presence 
of jets is a result of a significant increase of the accretion 
rate through the disk, so that an excess of angular momentum 
of the accreting material has to be removed via jets. 
A fraction of the disk material, accreted onto the WD, enhances 
nuclear shell burning on its surface. As a consequence, the 
high radiation output makes the surrounding disk unstable 
to warping and enhances the stellar wind during outbursts 
(see Sect.~\ref{ss:aap} for some details). 
The rest of the accreting material is liberated in the form of 
jets, whose kinetic energy comes from their accretion energy. 
This leads to a gradual removal of the disk material from its 
inner parts onto the WD and into the jets. As a result, the gradual 
decrease of fueling the WD and the release of the accretion energy 
lead to a gradual decline of the star's brightness, shrinking of 
the disk radius and vanishing of the jets as documented by 
observations in Fig.~\ref{fig:djc}. 

In a nutshell, expanding of the outer parts of the disk and 
sucking its inner parts by the luminous WD via the 
radiation-induced warping lead to a gradual dilution of the 
disk. This gives reasons for the observed disk-jets connection 
and explains why the launching of jets is transient. 
\subsection{On jets and Z~And-type outbursts}
\label{ss:aap}
During quiescent phases of symbiotic stars, no jets have been 
indicated, although their WDs accrete throughout the accretion disk 
and, in most cases, at high rates of $10^{-8} - 10^{-7}$\myr\ 
to power their high luminosities by stable hydrogen burning on 
the WD surface \citep[e.g.,][]{paczyt78,hach+96,shen+07}. 
To date, jets were indicated only during outbursts 
(see Sect.~\ref{s:intro}), during which a large optically thick 
disk is created around the WD \citep[][]{sk05}. The presence of 
jets thus confirms the presence of the disk and constrains accretion 
at a high rate, although enhanced mass-outflow from the hot 
component is directly indicated (see below). 

For the nuclearly powered symbiotics, the appearance of 
transient 
jets signals transient accretion at rates above the limit of 
the stable hydrogen burning on the WD surface. According to 
theoretical predictions, at these rates the luminosity of 
the burning WD can increase to the Eddington limit 
\citep[e.g., Fig.~2 of][]{shen+07}, and the mass-outflow from 
the WD in the form of wind enhances \citep[][]{hach+96}. 
Both characteristics are well supported by observations. 
(i) The high luminosity of the WD, burning hydrogen above 
the stable limit, was recently evidenced for AG~Peg during 
its 2015 Z~And-type outburst \citep[][]{sk+17}. For Z~And, 
the WD's luminosity of the order of $\sim 10^{37}$\es\ was 
determined for the 2000-03, 2006 and 2009-10 outbursts 
(see the point (ii) of Sect.~\ref{ss:warping}). 
(ii) Enhanced mass-outflow via the wind is directly indicated by 
broadening of emission lines during outbursts \citep[e.g.][]{fc+95}. 
Modeling the broad \ha\ wings, \cite{sk06} found an increase of 
the mass-loss rate via the wind from hot components during 
outbursts to $\ga 10^{-6}$\myr, i.e., factor of $\ga10$ higher 
with respect to values from quiescent phases. 

Appearance of jets during Z~And-type outbursts thus manifests 
their nature by nuclear burning of hydrogen on the WD surface 
at rates above the upper limit of the stable burning. 
The highly energetic events we observe during these outbursts 
(the luminosity close to the Eddington value and the wind 
outflow at $\gtrsim10^{-6}$\myr) require a relevant source 
of material that effectively fuels the WD. 
Indication of a large neutral disk at/around the orbital plane 
during Z~And-type outbursts was described in the point (i) of 
Sect.~\ref{ss:warping}. 
According to \cite{cask12} such a disk can be formed during 
outbursts by compression of the enhanced wind toward the 
equatorial plane due to rotation of the WD. This means 
that the disk consists of the material previously accreted 
onto the WD from the giant. Due to the thermal warping, this 
material is reaccreted again, which prolongs the period with 
a high luminosity. 
The continuous flow of material from the giant's wind helps 
to refill in the disk until the next possible instability will 
cause another (out)burst. This process can repeat up to 
depletion of the disk formed during outbursts. 
Then an accretion disk will be created from the giant's wind, 
and the system enter a quiescent phase. 

According to \cite{l+f08}, the main active phases of Z~And last 
approximately of 15--20 years and appear quasi-periodically 
with a separation of their centers by $\approx 20$ years 
(see their Fig.~1). Assuming this evolution also for the current 
active phase, it should cease around 2020. 
%
%----------------------------------------------------------------
%
\section{Summary}
\label{s:sum}
We continued monitoring of the prototypical symbiotic star 
Z~And after its major 2006 outburst, during which two-sided 
jets in the optical spectrum were indicated for the first time. 
We used the high-resolution spectroscopy around \ha, multicolor 
$UBVR_{\rm C}$ photometry, and high-time-resolution photometry 
to search for the reappearance of jets during the following 
outbursts of the current active phase (Fig.~\ref{fig:lc}). 
Our findings may be summarized as follows. 
\begin{enumerate}
\item
The bipolar jets reappeared during the major 2009-10 outburst, 
as indicated by well-pronounced S$^{-}$ and S$^{+}$ satellite 
components to the \ha\ line. During the smaller 2008 burst, 
only a single S$^{+}$ component was present 
(Sect.~\ref{ss:ha}, Figs.~\ref{fig:hba08} and \ref{fig:haev}). 
\item
The evolution of jets during both the 2006 and the 2009-10 outbursts 
was similar. The jets appeared around the outburst maxima, 
weakened along the decline of the star's brightness, and ceased 
after $\approx$5 and $\approx$10 months during 2006 and 2009-10 
outbursts. A large variation in jets radial velocities was 
measured at their emergence, but after 1--3 months they settled 
on a constant level of $\sim\pm1100$ and $\sim\pm1750$\kms, 
respectively (Sect.~\ref{ss:comp}, Fig.~\ref{fig:rvs}). 
\item
During both outbursts, a smooth light variation within 
$\Delta B \sim 0.06$\,mag on the timescale of hours developed 
from rapid, $\lesssim$0.02\,mag, stochastic fluctuations outside 
the outbursts 
(Sects.~\ref{ss:photevol} and \ref{ss:comp}, Fig.~\ref{fig:warp}). 
This type of variability is produced by the outer rim of the 
flared disk that develops during outbursts 
(see point (iii) of Sect.~\ref{ss:warping}). 
\item
Ejection of jets with simultaneous emergence of the 
hours-timescale photometric variation and the measured 
disk-jets connection can be caused by the radiation-induced 
warping of the inner disk due to the outburst of the central 
WD (Sects.~\ref{ss:warping}, \ref{ss:rvs} and \ref{ss:djc}). 
\item
Jets launched by nuclearly powered symbiotics require 
transient accretion at rates above the upper limit of the 
stable hydrogen burning on the WD surface. 
The jets thus prove the nature of Z~And-type outbursts by 
their ignition at these rates (Sect.~\ref{ss:aap}). 
Then the WD generates the luminosity of the order of 
$10^4$\lo\ and drives the wind at $\ga10^{-6}$\myr. 
\item
The large disk created at the beginning of the outbursts 
($R_{\rm D}\sim 10-20$\ro, Fig.~\ref{fig:djc}) consists of 
the material originally accreted onto the WD from the giant 
\citep[][]{cask12}. 
It represents a reservoir of mass for fueling the burning WD, 
because its inner part can be reaccreted via the 
radiation-induced warping. In this way, a high luminosity of 
the burning WD can be sustained for a longer time, until 
depletion of the disk (Sect.~\ref{ss:aap}). 
The jets thus signal the presence of the reaccretion process, 
providing a link to long-lasting active phases of Z~And. 
\end{enumerate}
Observations of transient jets along the decline from the optical 
maximum of some Z~And-type outbursts (to-date observed for 
Hen~3-1341 \citep[][]{munari+05}, 
Z~And \citep[e.g.][]{burmeister+07} and 
BF~Cyg \citep[][]{sk+13})
represent new challenges for the theoretical modeling of  
the radiation-induced warping of disks formed during outbursts 
of symbiotic stars. This should provide us with a better 
understanding of the accretion process during nuclearly 
powered eruptions on the surface of WDs with prolonged 
stages of high luminosity and signatures of collimated 
mass ejection. 

\acknowledgments
The authors thank the anonymous referee for critical, but 
encouraging, comments on the original version of the manuscript. 
Theodor Pribulla is thanked for acquisition of spectra 
at the David Dunlap Observatory. Miloslav Tlamicha, Tereza 
Krej\v{c}ov\'a, Pavel Chadima, Lenka Kotkov\'a, Luk\'a\v{c} 
Pilar\v{c}\'{i}k, Jan Sloup, Kate\v{r}ina Ho\v{n}kov\'a, 
Jakub Jury\v{s}ek, Lud\v{e}k \v{R}ezba and Jan Fuchs, 
are thanked for their assistance in the acquisition of 
the spectra at the Ond\v{r}ejov observatory. 
We also acknowledge the variable-star observations from the AAVSO 
International Database contributed by observers worldwide and used
in this research. 
This work was supported by the Czech Science Foundation, grants
P209/10/0715 and GA15-02112S, by the Slovak Research and 
Development Agency under the contract No. APVV-15-0458, by 
the Slovak Academy of Sciences grant VEGA No. 2/0008/17 
and by the realization of the project ITMS No.~26220120029, 
based on the supporting operational Research and development 
program financed from the European Regional Development Fund. 
\appendix
\label{app}

\section{Dynamical timescale of the disk}
\label{app1}
Here we compare the timescale of the observed light variability 
(Sect.~\ref{ss:photevol}) with the dynamical time of the disk 
with radius $R_{\rm D}$ around the WD with the mass of 
$M_{\rm WD}$, 
\begin{equation}
 t_{\rm dyn} = 0.44\times
         \left(\frac{R_{\rm D}}{R_{\odot}}\right)^{2/3}
         \left(\frac{M_{\rm WD}}{M_{\odot}}\right)^{-1/2}
         ~{\rm hr} .
\label{eq:tdyn}
\end{equation}
A representative value of $R_{\rm D} = 10\,R_{\odot}$ 
(see Fig.~\ref{fig:djc}) and $M_{\rm WD} = 1\,M_{\odot}$ 
corresponds to $t_{\rm dyn} \sim 14$\,hrs. If we consider 
$R_{\rm D} \sim 5\,R_{\odot}$ for the outer part of the 
disk, then $t_{\rm dyn} \sim 5$\,hrs--well comparable 
with the variation in the optical light as measured during 
the outbursts with jets. On the other hand, shorter variations 
from a low state, $\lesssim 1$\,hr (Fig.~\ref{fig:warp}), 
correspond to $R_{\rm D}\lesssim 1.5\,R_{\odot}$. 
\section{Diffusion timescale of the disk}
\label{app2}
A contribution to the smooth light variation could also come 
from the inner warping part of the disk, whose flickering light 
is hidden, but can be reprocessed in the outer disk. The 
timescale for this process is given by the diffusion timescale, 
\begin{equation}
      t_{\rm dif} = R_{\rm out}^2/c\,l , 
\end{equation}
where $R_{\rm out}$ is the radius of the reprocessing medium, 
$c$ is the light speed, and $l$ is the mean-free path of 
a photon. 
For the opacity $\kappa_\lambda$ and density $\rho$ of the 
medium, $l = 1/\kappa_\lambda \rho$. If we can assume that 
$\rho$ and $\kappa_\lambda$ are constant along the path of 
the photon, then the corresponding optical depth 
$\tau_\lambda = \kappa_\lambda \rho R_{\rm out} = R_{\rm out}/l$, 
and the diffusion timescale can be expressed as 
$R_{\rm out}\tau_\lambda/c$, or 
\begin{equation}
   t_{\rm dif} = 6.45\times 10^{-4}\,
                 \frac{R_{\rm out}}{R_{\odot}} \tau_\lambda 
                 ~~{\rm hr} .
\label{eq:tdif}
\end{equation}
If the thickness $R_{\rm out} \sim 5$\ro\ and its total optical 
depth is as large as 1--10$^4$, the diffusion timescale 
$t_{\rm dif}$ = 12\,s--1.3\,days is sufficiently short to observe 
the inner disk light variation at its outer rim (which is the 
warm pseudophotosphere), well within the timescale of the jet 
ejection period ($\approx 1$\,years). 
%
%======================== References ========================
%

%
\end{document}